\documentclass[aps,prd,showpacs,eqsecnum,notitlepage,nofootinbib]{revtex4}
\usepackage[centertags]{amsmath}
\usepackage{amssymb}
\usepackage{amsfonts,mathrsfs}
\usepackage{hyperref}	
\usepackage{graphicx}
\usepackage{dcolumn}
\usepackage{bm}
\usepackage{tabularx}
\usepackage{array}
\usepackage{float}

\usepackage{color}
\newcommand{\blu}{{}}

\newcommand{\N}{\mathbb{N}}
\newcommand{\T}{\mathbb{T}}

\newcommand{\be}{\begin{equation}}
\newcommand{\ee}{\end{equation}}
\newcommand{\bea}{\begin{eqnarray}}
\newcommand{\eea}{\end{eqnarray}}
\newcommand{\bse}{\begin{subequations}}
\newcommand{\ese}{\end{subequations}}

\begin{document}

\title{Quantum channels from reflections on moving mirrors}

\author{Giulio Gianfelici$^1$ and Stefano Mancini$^{1,2}$}
\affiliation{$^1$School of Science and Technology, University of Camerino, I-62032 Camerino, Italy}
\affiliation{$^2$INFN Sezione di Perugia, I-06123 Perugia, Italy}

\date{\today}

\begin{abstract}
Light reflection on a mirror can be thought as a simple physical effect. However if this happens when the mirror moves a rich scenario opens up. Here we aim at analyzing it from a quantum communication perspective.
In particular, we study the kind of quantum channel that arises from (Gaussian) light reflection upon an accelerating mirror. Two competing mechanisms emerge in such a context, namely photons production by the mirror's motion and {\blu interference between modes}.
As consequence we find out a quantum amplifier channel and quantum lossy channel respectively below and above a threshold frequency (that depends on parameters determining mirror's acceleration).
Exactly at the threshold frequency the channel behaves like a purely classical additive channel,
while it becomes purely erasure for large frequencies.
In addition the time behavior of the channel is analyzed by employing wave packets expansion of the light field.
\end{abstract}


\maketitle

\section{Introduction}
\label{sec:Intro}

The subject of quantum channels is becoming central in theoretical and mathematical physics (see Ref.\cite{Caruso:2014} for a recent review). Formally a quantum channel is a linear completely positive and trace preserving map acting on the set of states (density operators) living in a Hilbert space. Since any physical process involves a state change, it can be regarded as a quantum channel mapping the initial state to the final state. As such it can be characterized in terms of its information transmission capability.
Among quantum channels the Gaussian ones play a dominant role because they are easy to handle from a formal point of view and also because they are easily implemented with quantum electromagnetic fields \cite{Weedbrook:2012}.
In such optical quantum communication context a mirror turns out to be a fundamental device. It is however always considered at rest and hence causing a phase shift on the light field impinging on it. Considering it as a moving object we can figure out a frequency shift effect whenever it moves at constant velocity (Doppler shift). Nevertheless the situation becomes more involves if it accelerates,
 because then particles (photons) can be produced.
Actually in the relativistic framework a simple theoretical manifestation of particles production is
provided by the moving mirror model of DeWitt~\cite{DeWitt:1975}, and Davies and
Fulling~\cite{Davies:1976,Davies:1977} describing the
disturbance of a field by an accelerated boundary.
As such model matured~\cite{CD:1977,FV:1982,UW:1982,Carlitz:1986}, it became evident that accelerating boundaries could be used to understand entropy production \cite{Mukohyama:2000,Page:2009},
the relationship between particles and energy \cite{Walker:1984}, and
thermodynamical paradoxes \cite{Davies:1982,Walker:1985,Helfer:2000}.

Recently the notion of quantum channel has been extended into general relativistic settings
\cite{BHP:2009,OBW:2011,BHP:2012,MHM:2012,L:2016}.
One pursued idea was to consider the particles creation due to the expansion of the universe as
generating noisy effects once focusing on a single mode and tracing away all the other \cite{Mancini:2014}.
Following this line of reasoning we aim here at determining quantum channels arising
from reflection of a scalar field upon a moving boundary. In particular we shall employ an
accelerating mirror model in (1+1) dimensions and considering one of the modes where the field is expanded as encoding the signal while the other modes as environment,
we shall show that on the one hand information tends to be washed out by {\blu interference between modes} and on the other hand tends to be reinforced by photons production.
Given this two competing mechanisms,
the channel turns out to be Gaussian, specifically
quantum amplifier or a purely classical additive or quantum lossy depending on the considered regime.
Since expansion of of the field over plane waves gives rise to divergences
we shall also consider wave packets, which allows us to analyze time behavior of the channel.

We start considering a quantized massless
scalar field $\Phi$ in flat spacetime (1+1 dimensions) that obeys boundary conditions on
a perfectly reflecting mirror (throughout this paper units are used such that $\hbar = c = 1$).
The field satisfies the wave equation
\be
\Box \Phi = 0.
\ee
As usual we will expand the field in terms of mode functions that we consider parameterized by the frequency $\omega$.  Denoting them as $\phi_\omega$, they obey the wave equation
\be
\left(- \partial_t^2 + \partial_x^2 \right) \phi_\omega= 0,
\ee
or equivalently
\be
-\, \partial_u \partial_v \phi_\omega = 0,
\ee
where
\be
u \equiv t - x, \quad
v \equiv t + x.
\label{uv}
\ee
Here we shall consider solutions to the mode equation
that lie to the right of the mirror and include reflection from the
mirror's surface (see Methods \ref{sec:quantiz}).
We shall also consider mirror trajectories that begin at
past timelike infinity, $i^-$.
In this case past null infinity, $\mathscr{I}^-$, just consists of the surface $u = - \infty$
which results a Cauchy surface.  Furthermore, if the mirror trajectory ends at future timelike infinity,
$i^+$, then future null infinity, $\mathscr{I}^+$, just consists of the
surface $v =  \infty$ which results a Cauchy surface as well.  However, if the trajectory
is asymptotic to a null ray $v = v_0$, then $\mathscr{I}^+$ has two parts,
$\mathscr{I}^+_R$ and $\mathscr{I}^+_L$ \cite{Carlitz:1986}.
The former lies at $v = \infty$, while the latter consists of the part of the surface
$u = \infty$ which goes from $v = v_0$ to $v = \infty$.  Taken together they
provide a Cauchy surface.

In presence of the mirror we can consider either mode functions that
are positive frequency at $\mathscr{I}^-$ (corresponding to the
\textit{in} vacuum state), or mode functions that are positive frequency
at $\mathscr{I}^+$ (correspond to the \textit{out} vacuum state).

We employ a specific trajectory, namely the Carlitz and Willey trajectory~\cite{Carlitz:1986},
\be
z(t) = -t - \frac{1}{\kappa}W(e^{-2\kappa t}),
\label{cw-trajectory}
\ee
meaning that the mirror is at the position $x = z(t)$ at time $t$.
Here $\kappa\in\mathbb{R}$ is a free parameter and $W$ stands for the Lambert $W$ function.
The choice of this trajectory is motivated by the easiness to handle it compared with the few others
that admit analytical expressions \cite{GAE:2013} together with the fact that it shows an horizon.

Since the positive frequency modes at $\mathscr{I}^{-}$, $\phi^{in}_{\omega'}$ form a complete
set, we can expand modes at $\mathscr{I}^{+}$ in terms of them,
\be
\phi^J_\omega = \int_0^\infty d \omega' \left[ \alpha^J_{\omega \omega'}
\phi^{ in}_{\omega'} + \beta^J_{\omega \omega'}
\phi^{ in \;*}_{\omega'} \right],
\label{phiJomega}
\ee
with $J$ representing either $R$ or $L$ and
$\alpha^J_{\omega \omega'}$, $\beta^J_{\omega \omega'}$ the Bogolubov coefficients (see Methods \ref{sec:quantiz}).
The ladder operators in different regions will be related by (see Methods \ref{sec:quantiz})
\bse
\bea
b^J_\omega &=& \int_0^\infty d \omega'
\left[ (\alpha^J_{\omega \omega'})^* a^{ in}_{\omega'}
- (\beta^J_{\omega \omega'})^* a^{{ in} \; \dagger}_{\omega'} \right],\\
\left(b^J_\omega\right)^\dag &=& \int_0^\infty d \omega'
\left[ \alpha^J_{\omega \omega'} a^{{ in}\; \dag}_{\omega'}
- \beta^J_{\omega \omega'} a^{{ in}}_{\omega'} \right].
\eea
\label{abBog}
\ese
In Ref.~\cite{Carlitz:1986} analytic expressions were found for the
Bogolubov coefficients, namely
\bse
\bea
\alpha^R_{\omega\omega'} &=&
\frac{1}{4\pi\sqrt{\omega\omega'}}
\left\{-\frac{2\omega}{\kappa}e^{-\pi\omega/2\kappa}
\left(\frac{-\omega'}{\kappa}\right)^{-i\omega/\kappa}
\Gamma\left[\frac{i\omega}{\kappa}\right]\right\},\\
\beta^R_{\omega\omega'} &=&
\frac{1}{4\pi\sqrt{\omega\omega'}}
\left\{-\frac{2\omega}{\kappa}e^{-\pi\omega/2\kappa}
\left(\frac{\omega'}{\kappa}\right)^{-i\omega/\kappa}
\Gamma\left[\frac{i\omega}{\kappa}\right]\right\}.
\eea
\label{beta-c-w}
\ese


\section{Results}

\subsection{Quantum channel from plane waves reflection}
\label{sec:planewave}

We now consider all modes of $\mathscr{I}^-$ as encompassing signal and environments modes of a usual Stinespring dilation of a quantum channel, then transformed
into reflected modes in $\mathscr{I}^+_R$ through the unitary associated to
Bogoliubov transformation \eqref{abBog}.

The annihilation and creation operators of modes in $\mathscr{I}^-$ satisfy the following commutation relation
\be
\left[a^{{ in}}_{\omega} , a^{{ in}\, \dag}_{\omega'} \right]=\delta(\omega-\omega'),\qquad
\left[a^{{ in}}_{\omega} , a^{{ in}}_{\omega'} \right]=0.
\ee
Suppose that $\omega$ is the frequency of the signal mode, then
$a^{in}_\omega$ and $a^{in\,\dag}_{\omega}$ are the ladder operators of the input and $b^R_\omega$ and $(b^R_{\omega})^\dag$ those of the output.
They are related to the input ladder operators by the Bogoliubov transformation \eqref{abBog}.

Then, defining the quadrature operators
\bse
\bea
Q_{in}(\omega)&\equiv&\frac{a_\omega^{ in}+a_\omega^{{ in}\,\dag}}{\sqrt{2}},
\qquad
Q_{out}(\omega)\equiv\frac{b^R_\omega+(b_\omega^R)^\dag}{\sqrt{2}},\\
P_{in}(\omega)&\equiv&\frac{a^{ in}_\omega-a_\omega^{{ in}\,\dag}}{i\sqrt{2}},
\qquad\;
P_{out}(\omega)\equiv\frac{b^R_\omega-(b^R_\omega)^\dag}{i\sqrt{2}}.
\label{QPdef}
\eea
\ese
we have the input and output covariance matrices defined as
\be
V_{in}\equiv\begin{pmatrix}
\langle Q_{in}^2 \rangle & \frac{\langle Q_{in}P_{in}+P_{in}Q_{in}\rangle}{2} \\
 \frac{\langle Q_{in}P_{in}+P_{in}Q_{in}\rangle}{2} & \langle P_{in}^2 \rangle
\end{pmatrix},
\quad
V_{out}\equiv\begin{pmatrix}
\langle Q_{out}^2 \rangle & \frac{\langle Q_{out}P_{out}+P_{out}Q_{out}\rangle}{2} \\
 \frac{\langle Q_{out}P_{out}+P_{out}Q_{out}\rangle}{2} & \langle P_{out}^2 \rangle
\end{pmatrix},
\ee
where we implicitly assumed zero first moments.

$V_{in}$ and $V_{out}$ can be related (see Methods \ref{sec:planewave}) by
\be
V_{out}=\T V_{in} \T^T+\N.
\label{VoutVin}
\ee
where
\be
\label{contTN}
\T= S_{\omega\omega}\:, \qquad\qquad \N=-\frac{1}2\,S_{\omega\omega}\,S_{\omega\omega}^{T}+ \frac{1}2\int\,d\omega'\,S_{\omega\omega'}\, S_{\omega\omega'}^{T}.
\ee
being
\be
S_{\omega\omega'}=
\begin{pmatrix}
\Re\left(\alpha^R_{\omega,\omega'}-\beta^R_{\omega,\omega'}\right)
& \Im\left(\alpha^R_{\omega,\omega'}+\beta^R_{\omega,\omega'}\right) \\
-\Im\left(\alpha^R_{\omega,\omega'}-\beta^R_{\omega,\omega'}\right)
& \Re\left(\alpha^R_{\omega,\omega'}+\beta^R_{\omega,\omega'}\right)
\end{pmatrix},
\ee
a symplectic matrix.

Using Eq.\eqref{beta-c-w} for the Carlitz-Willey model, we can easily arrive at
\be
S_{\omega\omega'}=
\frac{1}{\pi\kappa}\sqrt{\frac{\omega}{\omega'}}\, \left|\Gamma\left(\frac{i\omega}{\kappa}\right)\right|\,
\begin{pmatrix}
\cosh\left(\frac{\pi\omega}{2\kappa}\right)\,\cos\theta_{\omega\omega'} &-\cosh\left(\frac{\pi\omega}{2\kappa}\right)\,\sin\theta_{\omega\omega'}\\
\sinh\left(\frac{\pi\omega}{2\kappa}\right)\,\sin\theta_{\omega\omega'}& \sinh\left(\frac{\pi\omega}{2\kappa}\right)\,\cos\theta_{\omega\omega'}
\end{pmatrix},
\ee
where $\theta_{\omega\omega'}\equiv\frac{\omega}{\kappa}\, \ln\left(\frac{\omega'}{\kappa}\right)- \arg\left[\Gamma\left(\frac{i\omega}{\kappa}\right)\right]$. Hence
\be
\label{CWT}
\mathbb{T}=\frac{1}{\pi\kappa}\, \left|\Gamma\left(\frac{i\omega}{\kappa}\right)\right|\,
\begin{pmatrix}
\cosh\left(\frac{\pi\omega}{2\kappa}\right)\,\cos\theta_{\omega\omega} &-\cosh\left(\frac{\pi\omega}{2\kappa}\right)\,\sin\theta_{\omega\omega}\\
\sinh\left(\frac{\pi\omega}{2\kappa}\right)\,\sin\theta_{\omega\omega}& \sinh\left(\frac{\pi\omega}{2\kappa}\right)\,\cos\theta_{\omega\omega}
\end{pmatrix}.
\ee
The matrix $S_{\omega\omega'}S_{\omega\omega'}^{T}$ reads
\be
\label{CWdY}
\begin{split}
S_{\omega\omega'}S_{\omega\omega'}^{\rm T}= & \frac{1}{2\pi\kappa\omega'}
\begin{pmatrix}
\coth\left(\frac{\pi\omega}{2\kappa}\right)  & 0 \\
0 & \tanh\left(\frac{\pi\omega}{2\kappa}\right)
\end{pmatrix}.
\end{split}
\ee
The integrals of its entries over the frequency $\omega'$ diverge in both infrared and ultraviolet limit. Thus we consider two cutoff, $\Omega_0$ and $\Omega_\infty$ to have a meaningful expression
for $\N$, that is
\be
\label{CWN}
\mathbb{N}=\frac{1}{4\pi\kappa}\left(\ln\left(\frac{\Omega_\infty}{\Omega_0} \right)-\frac{1}{\omega}\right)\,
\begin{pmatrix}
\coth\left(\frac{\pi\omega}{2\kappa}\right) & 0 \\
0 & \tanh\left(\frac{\pi\omega}{2\kappa}\right)
\end{pmatrix}.
\ee
Eq. \eqref{VoutVin} represents a Gaussian channel
and given the form of  $\mathbb{T}$ and $\mathbb{N}$ we can conclude (see Methods \ref{sec:planewave}) that such a channel is characterized by the following two parameters
\be
\label{CWtau}
\tau=\frac{1}{2\pi\omega\kappa}, \qquad\qquad\bar{n}=
\begin{cases}
\frac{\omega\ln(\Omega_\infty/\Omega_0)+2\pi\omega\kappa-2}{2- 4\pi\omega\kappa},\quad\text{for }\tau<1, \\ \\
\frac{\omega\ln(\Omega_\infty/\Omega_0)-2\pi\omega\kappa}{4\pi \omega\kappa-2},\quad\text{for }\tau>1, \\ \\
\frac{1}{4\pi\kappa}\,\ln(\Omega_\infty/\Omega_0)-\frac{1}{2},
\quad\text{for }\tau=1.
\end{cases}
\ee
{\blu The parameter $\tau$ whenever $<1$ (resp.$ >1$) can be interpreted as the transmissivity of a beam  splitter (resp. as the gain of a parametric amplifier) through which the radiation mode
$\omega$ pass. Furthermore the parameter $\bar{n}$ represents the number of thermal photons added to that mode. Then, according to Gaussian channels classification (see Methods \ref{sec:planewave}),} if $\tau<1$ (i.e. $\omega>1/2\pi\kappa$) we have a lossy channel, while for $\tau>1$ (i.e. $\omega<1/2\pi\kappa$) a linear amplifier. Instead, for $\tau=1$ (i.e. $\omega=1/2\pi\kappa$) we recover a classical additive noise channel.
These different regimes appear because of the presence of two competing mechanisms:

i) particles creation due to the mirror's motion that leads to amplification of the signal, but also to \emph{added noise}. This effect prevails at low frequency because of the easiness of exciting low frequency modes and it is somehow a manifestation of Dynamical Casimir effect \cite{DCE1,DCE2}.

ii) {\blu interference between modes (for the Doppler shift due to the reflection of modes on the mirror)} that causes losses photons from the signal mode to other (environment) modes. This effect preveals at high frequency where the other one is negligible.

The expression \eqref{CWtau} for $\tau$ diverges for $\omega=0$.
The result is not surprising, since one cannot measure particles of infinite wavelength and their eventual amplification. In the limit of $\omega\rightarrow\infty$ the loss of photons for {\blu interference between modes} is complete and the signal is completely erased.

The presence of infrared and ultraviolet cutoffs in the computation of the matrix $\N$ might appear as an artifact, hence to avoid them we will next employ wave packets \cite{Hawking:1975} rather than plane waves. This approach allows us to also study
time-dependent aspects.


\subsection{Quantum channel from wave packets reflection}
\label{sec:wavepack}

A wave packet can be constructed by integrating the field mode $\phi_\omega$
over a finite range of frequencies with a particular weighting function, e.g.
\be
\phi_{jn} \equiv \frac{1}{\sqrt{\epsilon}}\int_{j\epsilon}^{(j+1)\epsilon} d\omega\;
e^{2\pi i \omega n/\epsilon} \phi_\omega,
\label{mode-packet}
\ee
with $n$ taking on integer values and $j$ taking on nonnegative integer values.
Actually the value of $j$ is related to the frequency
of the modes in the packet, in fact $(j+1/2)\epsilon$ gives the frequency at
the center of the range and $\epsilon$ gives the width of the range.

The Bogolubov coefficients that
correspond to the wave packets  can be obtained directly from the
coefficients $\beta^R_{\omega\omega'}$ by using the same weighting, integrating
over frequency, and swapping the order of integration
\be
\beta^{R}_{j n, \omega'} =
\frac{1}{\sqrt{\epsilon}}\int_{j\epsilon}^{(j+1)\epsilon}
d\omega \; e^{2\pi i \omega n/\epsilon} \beta^{R}_{\omega\omega'}.
\label{beta-packet}
\ee
From Eq. \eqref{beta-packet} with $\beta_{\omega,\omega'}^R$ given by Eq. \eqref{beta-c-w} we obtain
\be
\label{CWpack}
\beta^R_{jn,\omega'}=\frac{1}{2\pi\kappa\sqrt{\varepsilon\,\omega'}} \,\int_{j\varepsilon}^{(j+1)\varepsilon}\,d\omega\,e^{2\pi i\omega n/\varepsilon}\, \sqrt{\omega}\,e^{-\pi\omega/2\kappa}\, e^{-i\omega/\kappa \ln(\omega'/\kappa)}\, \Gamma\left[\frac{i\omega}{\kappa}\right],
\ee
and similarly
\be
\alpha^R_{jn,\omega'}=\frac{1}{2\pi\kappa\sqrt{\varepsilon\,\omega'}} \,\int_{j\varepsilon}^{(j+1)\varepsilon}\,d\omega\,e^{2\pi i\omega n/\varepsilon}\, \sqrt{\omega}\,e^{-\pi\omega/2\kappa}\, e^{-i\omega/\kappa \ln(-\omega'/\kappa)}\, \Gamma\left[\frac{i\omega}{\kappa}\right].
\ee
Then the output ladder operators are related to the input ones by the following Bogoliubov transformation that modify  those of Eq.\eqref{abBog}
\bse
\bea
b^R_{jn}&=&\int_0^\infty d\omega' \left[\alpha_{jn;\omega'}^* \, a^{in}_{\omega'}-\beta_{jn;\omega'}^*\, ( a_{\omega'}^{in} )^\dag \right],\\
(b^R_{jn})^\dag&=&\int_0^\infty d\omega' \left[ \alpha_{jn;\omega'}\, (a_{\omega'}^{in})^\dag
-\beta_{jn;\omega'}\, a^{in}_{\omega'} \right].
\eea
\ese
{\blu In practice with these we are considering as output the
signal reaching $\mathscr{I}^+_R$ in the frequency range $[j\epsilon, (j+1)\epsilon]$ and in the time range $\left[\frac{2\pi n-\pi}{\epsilon},\frac{2\pi n+\pi}{\epsilon}\right]$.
}
Defining the quadrature operators analogously to Eq.\eqref{QPdef}
\bse
\bea
Q_{in}(\omega)&\equiv&\frac{a_\omega^{ in}+a_\omega^{{ in}\,\dag}}{\sqrt{2}},
\qquad
Q_{out}(jn)\equiv\frac{b^R_{jn}+(b_{jn}^R)^\dag}{\sqrt{2}},\\
P_{in}(\omega)&\equiv&\frac{a^{ in}_\omega-a_\omega^{{ in}\,\dag}}{i\sqrt{2}},
\qquad\;
P_{out}(jn)\equiv\frac{b^R_{jn}-(b^R_{jn})^\dag}{i\sqrt{2}}.
\eea
\ese
we can relate $V_{in}$ and $V_{out}$ as (see Methods \ref{sec:wavepack})
\be
\label{VoutvsVinwp}
V_{out}=\T V_{in} \T^T+\N.
\ee
where now
\be
\T= S_{jn;\,\tilde{\omega}}\:, \qquad\qquad \N=-\frac{1}2\,S_{jn;\,\tilde{\omega}}\,S_{jn;\,\tilde{\omega}}^\mathrm{T}+ \frac{1}2\int\,d\omega'\,S_{jn;\,\omega'}\, S_{jn;\,\omega'}^\mathrm{T},
\ee
with
\be
S_{jn,\omega'}=
\begin{pmatrix}
\Re(\alpha_{jn,\omega'}^R-\beta_{jn,\omega'}^R) & \Im(\alpha_{jn,\omega'}^R+\beta_{jn,\omega'}^R) \\
-\Im(\alpha_{jn,\omega'}^R-\beta_{jn,\omega'}^R) & \Re(\alpha_{jn,\omega'}^R+\beta_{jn,\omega'}^R)
\end{pmatrix}.
\ee
and $\tilde{\omega}\equiv(j+1/2)\epsilon$.
The above symplectic matrix involves integrals like Eq. \eqref{CWpack}, which cannot be performed analytically. Hence the Bogoliubov coefficients
$\alpha_{jn,\omega'}$ and $\beta_{jn,\omega'}$ have been computed numerically for fixed values of
$\kappa$ and $\varepsilon$. By means of them we evaluated the matrices $\T$ and $\N$ and hence the parameters $\tau$ and $\bar{n}$ as functions of $n$ and $j$.
{\blu This allowed us to construct the evolution of the frequency distribution of channels in time, as it would be seen by a series of receivers spread out over a line of constant but large $v$.
}

\begin{figure}[h!]
  \centering
   \includegraphics[scale=0.55]{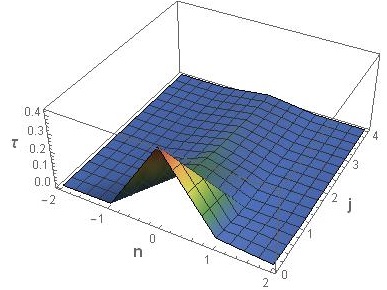}
   \hspace{1cm}
   \includegraphics[scale=0.55]{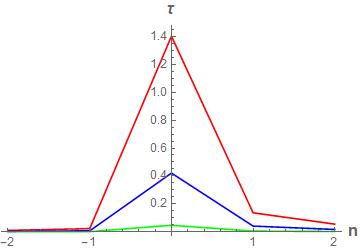}
  \caption{(Left) Parameter $\tau$ plotted vs $n$ and $j$ for $\kappa=1$ and $\varepsilon=0.1$.
  (Right) Parameter $\tau$ plotted vs $n$ for $j=0$ and different values of $\kappa,\varepsilon$ ($\kappa=0.3$ and $\varepsilon= 0.03$ red line; $\kappa=1$ and $\varepsilon=0.1$ blue line;
  $\kappa=9 $ and $\varepsilon=0.9$ green line).}
    \label{tau}
\end{figure}

In Fig.\ref{tau} (Left) the quantity $\tau$ is reported as function of $n$ and $j$ for a value of $\kappa$. The corresponding value of $\varepsilon$ has been chosen in order to have the highest value of the maximum of $\tau$. We can see a peak centered on $n=0$.
Moreover this peak takes the maximum value at $j=0$, because this value correspond to the lowest central frequency and it is the easiest to excite.
Indeed, the mirror create more low frequency photons and less high frequency photons.
It is worth remarking that the accelerated motion of the mirror produces (particles) radiation, which tends to amplify the signal. However as long as the central frequency $j$ of the wave packet increases the signal modes become harder to excite and hence the amplification effect diminishes,
while the loss due to {\blu interference between modes} increases. Hence we have a transition from $\tau>1$ to
$\tau <1$ by increasing $j$ (depending on the maximum value for $j=0$, we can still have $\tau>1$ for $j=1$, however for $j=2$ it is always $\tau <1$).
Another features concerns the optimal value of $\varepsilon$ which increases by increasing $\kappa$. This occurs because by increasing the acceleration many more modes are excited and hence the wave packet must be wider to benefits of amplification effect.

In Fig.\ref{tau} (Right) the quantity $\tau$ is reported as function of $n$ at $j=0$ for different values of $\kappa$. We can see that the peak is always centered in $n=0$
notwithstanding the particles production due to the mirror's motion results constant in time.
This is due to the fact that in Eq. \eqref{mode-packet}
we get a non-oscillatory quantity for $|n|\to 0$ and $j\to 0$, which reduces the
{\blu interference between modes} and maximize the amplification effect.
Actually the amplification effect ($\tau>1$) becomes manifest by increasing $\kappa$.
The value $\tau=1$ corresponding to a purely classical additive channel can be
attained for $n=j=0$ at $\kappa=0.4$.

\begin{figure}[h!]
  \centering
   \includegraphics[scale=0.55]{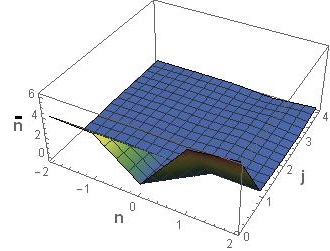}
   \hspace{1cm}
   \includegraphics[scale=0.55]{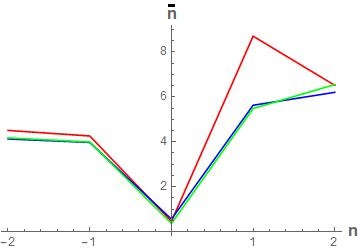}
  \caption{(Left) Parameter $\bar{n}$ plotted vs $n$ and $j$ for $\kappa=1$ and $\varepsilon=0.1$.
  (Right) Parameter $\bar{n}$ plotted vs $n$ for $j=0$ and different values of $\kappa,\varepsilon$
  ($\kappa=0.3$ and $\varepsilon= 0.03$ red line; $\kappa=1$ and $\varepsilon=0.1$ blue line;
  $\kappa=9 $ and $\varepsilon=0.9$ green line).}
    \label{barn}
\end{figure}

In Fig.\ref{barn} (Left) the quantity $\bar{n}$ is reported as function of $j$ and $n$ for the same values of $\kappa$ and $\varepsilon$ of Fig.\ref{tau} (Left).
We may notice that for $j\ge 1$ the number of added thermal photons reduces almost to zero.
A remarkable feature is that for $j=0$ the minimum of $\bar{n}$ is achieved for $n=0$.
This happens for all values of $\kappa$ as can be seen in Fig.\ref{barn} (Right).
By referring to the dashed line (corresponding to $\tau>1$ at $n=0$), the reduction of the number of added thermal photons below one, leads to an almost perfect quantum amplifier channel.


\section{Discussion}
\label{sec:conclusion}

We have investigated the kind of quantum channel that arises from light reflection upon an accelerating mirror following the Carlitz-Willey trajectory \cite{Carlitz:1986}.
We first approached the problem by considering plane waves impinging on the mirror.
It comes out the possibility of having a
quantum amplifier channel or a purely classical additive channel or a quantum lossy channel depending mainly on the frequency of the signal mode.
This should be ascribed to the presence of two competing mechanisms emerging in the considered model, namely photons production by the mirror's motion and {\blu interference between modes.}
However it was possible to compute the added noise only by introducing frequency cutoffs.
Then we considered a wave packets based approach and the same typologies of quantum channels come out without the need of frequency cutoffs.
Furthermore it results the interesting feature of realizing an almost perfect amplifier channel when the wave packet is centered on the lowest possible frequency.

Above all, with the wave-packets approach we got a time dependent picture.
Since the integer $n$ characterizes the time bin at the receiver,
we realized that a perfect time matching for decoding should be in order, not to erase all information.
This however is practically independent on the parameters entering into the Carlitz-Willey model.
Viewed in the other way around, the channel can be regarded as varying in time ($\tau$ and $\bar{n}$ as functions of $n$). Moving on from this example we could conceive a quantum channel varying from one to another use and call for an information theoretic characterization of this class of channels (widely studied in the classical setting \cite{Sutivong:2005}).

{\blu
The outlined approach can be interesting for technological applications like earth-satellite or satellite-satellite communication (see \cite{Wang:2012,Elser:2015,Val2015,Liao:2016} for seminal works on quantum key distribution and more recently \cite{Yin:2017,Ren:2017} for other protocols), considering the moving mirror as a (simplified) model of a satellite. In such a case however one should deal with a trajectory free from events horizon (unlike the Carlitz-Willey one). Unfortunately, even in 1+1 spacetime, there are just few trajectories that can be treated analytically \cite{GAE:2013}. Among them the so-called Darcx seems the most appropriate. It is described by
\be
z(t)=-\frac{\xi}{\nu}\sinh^{-1}\left(e^{\nu t}\right), \quad 0< |\xi|<1, \quad\nu\in\mathbb{R}.
\ee
For it the mirror begins at rest and is asymptotically inertial in the future, where it approaches a constant speed that is less than that of light. This means that there is a peak of acceleration around $t=0$. The width of such peak decreases by increasing $|\nu|$, while the maximum achieved value is approximately $|\xi\nu|$. Loosely speaking this might describe the launch and orbital setting of a satellite or the come into view and subsequent departure from a ground station of a satellite orbiting around earth.
For such a trajectory (actually for all trajectories of \cite{GAE:2013} having no events horizon) the plane wave approach leads to undefined $\tau$. However the wave packet approach gives valid results.

\begin{figure}[h!]
  \centering
   \includegraphics[scale=0.5]{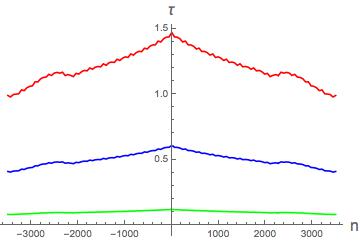}
   \hspace{1cm}
   \includegraphics[scale=0.5]{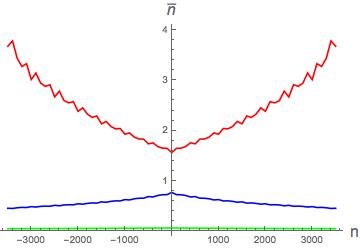}
  \caption{\blu
  Parameters $\tau$ (Left) and $\bar{n}$ (Right) plotted vs $n$ for $j=0$ and different values of $\xi$, $\nu$ such that $|\xi\nu|=10^{-50}$ (red lines $\xi=5.6\times 10^{-27}$, blue lines $\xi=3.6\times 10^{-27}$, and green lines $\xi=1.6\times 10^{-27}$). For all curves it is $\epsilon=2\times 10^{-44}$.}
    \label{taunbarDarcx}
\end{figure}

In Fig.\ref{taunbarDarcx} it is reported $\tau$ (on the Left) and $\bar{n}$ (on the Right) vs $n$ for the lowest frequency bin $j=0$, showing different channel's regimes (similarly to
Figs.\ref{tau},\ref{barn} (Right)). A realistic peak value of satellite acceleration ($5g$ \cite{Marsten:2014}) is employed, thus obtaining $|\xi\nu|=10^{-50}$. We can see that the amplification effect is washed out by the decrement of $|\xi|$ (and contextual increment of $|\nu|$). This time however there not exist an optimal value of $\epsilon$, but the smaller it is,
the larger $\tau$ results. Hence, the possibility to avoid losses (or even erasure) mostly relies on the accuracy of controlling the wave packet size (the value chosen in Fig.\ref{taunbarDarcx} corresponds approximately to a size of $1 Hz$). The smallness of $\epsilon$ also makes the plots extending over a much wider time window (range of values for $n$) with respect to
Figs.\ref{tau},\ref{barn}.
}

{\blu Finally, the presented study can also be interesting for a} relativistic quantum information theory given the correspondence between a black hole and an accelerating mirror \cite{Good:2016}.
Along this line we can foresee investigations in terms of quantum channels for field scattered by variation of the underlying spacetime.
In fact, the accelerating mirror plays the same role as a time-dependent background geometry, i.e. gravitational field, given that the mirror motion distorted the field modes.
The difference is that the distortion of the modes occurs suddenly (upon reflection) rather than gradually during an extended period of geometrical disruption, as in the gravitational case.


\section{Methods}

\subsection{Field quantization with a moving mirror}
\label{sec:quantiz}

Orthonormal mode functions satisfy
\be
(\phi_\omega(x),\phi_{\omega'}(x)) = - (\phi^{*}_\omega(x),\phi^{*}_{\omega'}(x))
= \delta(\omega-\omega'), \quad
(\phi_\omega(x), \phi^{*}_{\omega'}(x)) = 0,
\label{normalization}
\ee
where the scalar product is defined as
\be
(\phi_1,\phi_2) \equiv - i \int_\Sigma \; d \Sigma \, n^\mu
\left[\phi_1(x) \stackrel{\leftrightarrow}{\partial_\mu}\phi^*_2(x)\right].
\label{scalar-product}
\ee
Here $\Sigma$ is any Cauchy surface for the spacetime
and $n^\mu$ is a future-directed unit normal to that surface~\cite{Birrell:1982}.

In case of Minkowski spacetime with no boundaries, the following orthonormal modes
\be
\phi_{\omega u} = \frac{1}{\sqrt{4 \pi \omega}}e^{- i \omega u}, \quad
\phi_{\omega v} = \frac{1}{\sqrt{4 \pi \omega}}e^{- i \omega v},
\label{flat-modes}
\ee
are commonly chosen.
Then the field $\Phi$ is expanded as
\be
\Phi = \int_0^\infty \frac{d \omega}{\sqrt{4 \pi \omega}}
\left[ a_{\omega u} e^{- i \omega u}  + a_{\omega v} e^{- i \omega v} +
a^\dagger_{\omega u} e^{+ i \omega u} + a_{\omega v}^\dagger e^{+ i \omega v}
\right],
\ee
with $a_{\omega u}$, $a_{\omega v}$, $a^\dagger_{\omega u}$, and
$a^\dagger_{\omega v}$ annihilation and creation operators.

Eqs.\eqref{flat-modes} need to be modified in the presence of a mirror following
a trajectory $z(t)$ (meaning that the mirror is at the position $x = z(t)$ at time $t$),
so that the spacetime possesses a boundary.

In $\mathscr{I}^-$ the mode functions read
\be
\phi^{in}_{\omega'} = \frac{1}{\sqrt{4 \pi \omega'}}
\left[ e^{- i \omega' v} - e^{- i \omega' p(u)} \right].
\label{p-modes}
\ee
It must be $v = p(u)$ at the location of the mirror in order that
mode functions \eqref{p-modes}
vanish at the mirror.  Using \eqref{uv}, we find
\be
p(u) = t_m(u) + z(t_m(u)),
\label{pumeq}
\ee
where $t_m(u)=u-z$.

In a general left-right construction (for mirrors having a horizon at
$v_0$), there are two sets of mode functions that are positive frequency
at $\mathscr{I}^+$.  One, $\phi^R_\omega$, with functions
nonzero at $\mathscr{I}^+_R$ and zero at $\mathscr{I}^+_L$.
The other, $\phi^L_\omega$, with functions zero at $\mathscr{I}^+_R$ and nonzero
at $\mathscr{I}^+_L$.
The former are given by
\be
\phi_\omega^{R,\;out} = \frac{1}{\sqrt{4 \pi \omega}} \,
\left[ e^{-i \omega f(v)} - e^{-i \omega u} \right], \qquad v < v_0.
\label{f-modes}
\ee
For trajectories that begin at past timelike infinity, $i^-$, it is
$-\infty < u < \infty$. Furthermore mirrors that are asymptotically
inertial in the future have $v_0 = \infty$.
We must have $u = f(v)$ at the location of the mirror in order these modes vanish at the mirror.
Using \eqref{uv}, we find
\be
f(v) =\overline{t}_m(v) - z\left(\overline{t}_m(v)\right),
\ee
where $\overline{t}_m(v)=v-z$.
There are no other modes if $v_0 = \infty$.  However, if the mirror's trajectory
is asymptotic to the null surface $v = v_0$, then we must also include the
set of modes $\phi^{L}_\omega$ reaching $\mathscr{I}^+_L$ and never
interacting with the mirror.

For the Carlitz and Willey trajectory~\cite{Carlitz:1986},
the associated functions $p(u)$ and $f(v)$ result
\be
p(u) = - \frac{1}{\kappa} e^{- \kappa u}, \qquad f(v)=-\frac{1}{\kappa}\log(-\kappa v).
\label{p-carlitz-willey}
\ee
It is easy to see that $\dot{z}\rightarrow \pm 1$ in the limits $t \rightarrow \mp \infty$
and that $z < 0$ for all time.  Actually the mirror trajectory begins at past timelike
infinity, $i^{-}$, and at late times approaches $v  = 0$.  The proper acceleration,
\be
\frac{\ddot z}{(1-{\dot z}^2)^{3/2}}= -\frac{\kappa}{2\sqrt{W(e^{-2 \kappa t})}},
\label{thermalacc}
\ee
results not constant over the time.

Using Eqs.~\eqref{normalization} the Bogolubov coefficients of \eqref{phiJomega} result
\bse
\bea
\alpha^J_{\omega \omega'} &=& (\phi^J_\omega,\phi^{ in}_{\omega'}),
\label{alpha-def} \\
\beta^J_{\omega \omega'} &=& -(\phi^J_\omega,\phi^{{ in}\; *}_{\omega'}).
\label{beta-def}
\eea
\label{alpha-beta-def}
\ese
Since the field $\Phi$ can be expanded in either of two ways,
\bse
\bea
\Phi &=& \int_0^\infty d \omega' \left[a^{ in}_{\omega'}
\phi^{ in}_{\omega'} +  a^{ in \; \dagger}_{\omega'}
\phi^{ in \; *}_{\omega'} \right] \\
     &=&  \sum_{J=L,R}  \int_0^\infty d \omega \left[b^J_\omega \phi^J_\omega
+  b^{J \; \dagger}_\omega \phi^{J \; *}_\omega \right],
\eea
\ese
the ladder operators in different regions, using $b^J_\omega = (\Phi,\phi^J_\omega)$~\cite{Birrell:1982},
will be related by \eqref{abBog}.


\subsection{Quantum channel from plane waves reflection}
\label{sec:planewave}

In order to relate $V_{in}$ and $V_{out}$, let us compute the entries of $V_{out}$
\be
\begin{split}
\langle Q_{out}^2(\omega) \rangle&=\frac{1}{2} \langle\left(b^R_\omega+b_\omega^{R\,\dag}\right)
 \left(b^R_\omega+b_\omega^{R\,\dag}\right)\rangle\\
 &=\frac{1}{2}\int d\omega' d\omega'' \left(\alpha_{\omega,\omega'}^{R\,*}-\beta_{\omega,\omega'}^R\right)
 \left(\alpha_{\omega,\omega''}^{R\,*}-\beta_{\omega,\omega''}^R\right)
 \langle a^{in}_{\omega'} a^{in}_{\omega''} \rangle\\
 &+\frac{1}{2}\int d\omega' d\omega'' \left(\alpha_{\omega,\omega'}^{R\,*}-\beta_{\omega,\omega'}^R\right)
 \left(\alpha_{\omega,\omega''}^R-\beta_{\omega,\omega''}^{R\,*}\right) \langle a_{\omega'}^R a^{R\, \dag}_{\omega''} \rangle\\
 &+\frac{1}{2}\int d\omega' d\omega'' \left(\alpha_{\omega,\omega'}^R-\beta^{R\,*}_{\omega,\omega'}\right)
 \left(\alpha^{R\,*}_{\omega,\omega''}-\beta_{\omega,\omega''}^R\right) \langle a^{in\,\dag}_{\omega'} a^{in}_{\omega''} \rangle\\
 &+\frac{1}{2}\int d\omega' d\omega'' \left(\alpha_{\omega,\omega'}^R-\beta_{\omega,\omega'}^{R\,*}\right)
 \left(\alpha_{\omega,\omega''}^R-\beta_{\omega,\omega''}^{R\,*}\right) \langle a^{R\,\dag}_{\omega'} a^{R\,\dag}_{\omega''} \rangle.
\end{split}
\ee
In $\mathscr{I}^-$, without loss of generality, we may consider uncorrelated modes, hence
we may write
\bse
\bea
\langle a^{in}_{\omega'} a^{in}_{\omega''} \rangle&=M(\omega')\delta(\omega'-\omega''),\\
\langle a^{in\,\dag}_{\omega'} a^{in}_{\omega''} \rangle&=N(\omega')\delta(\omega'-\omega''),
\eea
\label{MNwp}
\ese
where $M(\omega')$ (resp. $N(\omega'))$ is a complex (resp. real) valued function.
Therefore
\be
\begin{split}
\langle Q_{out}^2(\omega) \rangle
 &=\frac{1}{2}\int d\omega'  \left(\alpha_{\omega,\omega'}^{R\,*}-\beta_{\omega,\omega'}^R\right)^2 M(\omega')
 +\frac{1}{2}\int d\omega' \left(\alpha_{\omega,\omega'}^R-\beta_{\omega,\omega'}^{R\,*}\right)^2
M^*(\omega')\\
& +\frac{1}{2}\int d\omega' \left|\alpha_{\omega,\omega'}^{R\,*}-\beta_{\omega,\omega'}^R\right|^2
 N(\omega')
+\frac{1}{2}\int d\omega' \left|\alpha_{\omega,\omega'}^{R\,*}-\beta_{\omega,\omega'}^R\right|^2
 \left(1+N(\omega')\right).
\end{split}
\ee
We can further assume that all modes, but the signal one,
are initially in the vacuum, i.e. write
\bse
\bea
M(\omega')&=M\delta(\omega'-\omega),\\
N(\omega')&=N\delta(\omega'-\omega),
\eea
\label{MNdelta}
\ese
where $M$ and $N$ determine the entries of $V_{in}$.

It follows that
\be
\langle Q_{out}^2(\omega) \rangle
 = \frac{1}{2}\left(\alpha_{\omega,\omega}^{R\,*}-\beta_{\omega,\omega}^R\right)^2 M
 +\frac{1}{2}\left(\alpha_{\omega,\omega}^R-\beta_{\omega,\omega}^{R\,*}\right)^2 M^*
 +\left|\alpha_{\omega,\omega}^{R\,*}-\beta_{\omega,\omega}^R\right|^2\, N
+\frac{1}{2}\,\int d\omega'  \left|\alpha_{\omega,\omega'}^{R\,*}-\beta_{\omega,\omega'}^R\right|^2.
\label{qout}
\ee
Analogously we can find
\be
\langle P_{out}^2(\omega) \rangle
 =-\frac{1}{2}\left(\alpha_{\omega,\omega}^{R\,*}+\beta_{\omega,\omega}^R\right)^2 M
 -\frac{1}{2}\left(\alpha_{\omega,\omega}^R+\beta_{\omega,\omega}^{R\,*}\right)^2 M^*
+\left|\,\alpha_{\omega,\omega}^{R\,*}+\beta_{\omega,\omega}^R\right|^2\,N
+\frac{1}{2}\,\int d\omega'  \left|\alpha_{\omega,\omega'}^{R\,*}+\beta_{\omega,\omega'}^R\right|^2,
\label{pout}
\ee
and
\be
\begin{split}
\frac{\langle Q_{out}P_{out}+P_{out}Q_{out}\rangle}{2}&
=\frac{1}{2i}\left(\alpha_{\omega,\omega}^{R\,*}-\beta_{\omega,\omega}^R\right)
\left(\alpha_{\omega,\omega}^{R\,*}+\beta_{\omega,\omega}^R\right)M
-\frac{1}{2i}\left(\alpha_{\omega,\omega}^R-\beta_{\omega,\omega}^{R\,*}\right)
\left(\alpha_{\omega,\omega}^R+\beta_{\omega,\omega}^{R\,*}\right)M^*\\
&+\frac{1}{2i}\left(\alpha_{\omega,\omega}^{R\,*}+\beta_{\omega,\omega}^R\right)
\left(\alpha_{\omega,\omega}^R-\beta_{\omega,\omega}^{R\,*}\right) \, N
-\frac{1}{2i}\left(\alpha_{\omega,\omega}^R+\beta_{\omega,\omega}^{R\,*}\right)
\left(\alpha_{\omega,\omega}^{R\,*}-\beta_{\omega,\omega}^R\right)\, N\\
&+\frac{1}{2}\,\Im \int d\omega' \left(\alpha_{\omega,\omega'}^{R\,*}+\beta_{\omega,\omega'}^R\right)
\left(\alpha_{\omega,\omega'}^R-\beta_{\omega,\omega'}^{R\,*}\right).
\end{split}
\label{qpout}
\ee
As said before $M$ and $N$ determine the entries of $V_{in}$. Exploiting the definitions \eqref{MNwp}, \eqref{MNdelta} of $M$ and $N$, we can write
\bse
\bea
\langle Q_{in}^2\rangle&=&\frac{1}{2}+N+\frac{1}{2}M+\frac{1}{2}M^*,\\
\langle P_{in}^2\rangle&=&\frac{1}{2}+N-\frac{1}{2}M-\frac{1}{2}M^*,\\
\frac{\langle Q_{in}P_{in}+P_{in}Q_{in}\rangle}{2}&=&\frac{1}{2i}M- \frac{1}{2i}M^*.
\eea
\ese
These relations can be reverted to get
\bse
\bea
M&=& \frac{1}{2}\left(\langle Q_{in}^2\rangle -\langle P_{in}^2\rangle \right)
+i \frac{\langle Q_{in}P_{in}+P_{in}Q_{in}\rangle}{2},\\
N&=& \frac{1}{2}\left(\langle Q_{in}^2\rangle +\langle P_{in}^2\rangle \right)-\frac{1}{2}.
\eea
\label{MN}
\ese
Substituting Eq.\eqref{MN} into Eqs.\eqref{qout}, \eqref{pout}, \eqref{qpout} we get a relation between the entries of $V_{out}$ and those of $V_{in}$ in the form of \eqref{VoutVin}.

Gaussian channels
can be classified according to equivalence up to Gaussian
unitary transformations (see e.g. \cite{Lupo:2011}).
For each class, one can pick up a representative
channel associated with a pair in the canonical form $(\mathbb{T}_c,\mathbb{N}_c)$.
Hence we introduce the following symplectic matrix
\be
S_B=
\begin{pmatrix}
\sqrt{\tanh\left(\frac{\pi\omega}{2\kappa}\right)} & 0 \\
0 & \sqrt{\coth\left(\frac{\pi\omega}{2\kappa}\right)}
\end{pmatrix},
\ee
to find from \eqref{CWN}
\be
\mathbb{N}_c=S_B\,\mathbb{N}\,S_B^{\rm T}=\frac{1}{4\pi\kappa}\left(\ln\left(\frac{\Omega_\infty}{\Omega_0} \right)-\frac{1}{\omega}\right)\,\mathbb{I}.
\ee
Analogously we introduce the symplectic matrix
\be
S_A=
\begin{pmatrix}
\cos\theta_{\omega\omega} & \sin\theta_{\omega\omega} \\
-\sin\theta_{\omega\omega} & \cos\theta_{\omega\omega}
\end{pmatrix},
\ee
to get from \eqref{CWT}
\be
\mathbb{T}_c= S_B\,\mathbb{T}\,S_A= \frac{1}{\sqrt{2\pi\omega\kappa}}\, \mathbb{I}.
\ee
In view of the form of $(\mathbb{T}_c,\mathbb{N}_c)$, setting $\tau= \frac{1}{\sqrt{2\pi\omega\kappa}}$, the Gaussian channel results (see e.g. \cite{Lupo:2011}) amplifier $\tau\in(1,\infty)$, classical additive $\tau=1$, attenuator $\tau\in(0,1)$ and erasure $\tau=0$.
The corresponding number of thermal photons $\bar{n}$ can be determined setting respectively
$\frac{1}{4\pi\kappa}\left(\ln\left(\frac{\Omega_\infty}{\Omega_0} \right)-\frac{1}{\omega}\right)=(\tau-1)(\bar{n}+1/2)$, $=\bar{n}$, $=(1-\tau)(\bar{n}+1/2)$ and $=(\bar{n}+1/2)$.


\subsection{Quantum channel from wave packets reflection}
\label{sec:wavepack}

We can relate $V_{in}$ and $V_{out}$, proceeding like in Sec.\ref{sec:planewave}. We get for example
\be
\begin{split}
\langle Q_{out}^2(jn) \rangle
 &=\frac{1}{2}\int d\omega'  \left(\alpha_{jn;\omega'}^{R\,*}-\beta_{jn;\omega'}^R\right)^2 M(\omega')
 +\frac{1}{2}\int d\omega' \left(\alpha_{jn;\omega'}^R-\beta_{jn;\omega'}^{R\,*}\right)^2
M^*(\omega')\\
 &+\frac{1}{2}\int d\omega' \left|\alpha_{jn;\omega'}^{R\,*}-\beta_{jn;\omega'}^R\right|^2
 \left(1+N(\omega')\right)
 +\frac{1}{2}\int d\omega' \left|\alpha_{jn;\omega'}^{R\,*}-\beta_{jn;\omega'}^R\right|^2
 N(\omega').
\end{split}
\ee
Now we assume that only the central mode of the wave packet is initially populated, while all the other modes are initially in the vacuum, i.e. write
\bse
\bea
M(\omega')&=M\,\delta\left(\omega'-\tilde{\omega}\right),\\
N(\omega')&=N\,\delta\left(\omega'-\tilde{\omega}\right),
\eea
\ese
where $M$ and $N$ determine the entries of $V_{in}$ and
\be
\tilde{\omega}=\left(j+\frac{1}{2}\right)\varepsilon,
\ee
is the central frequency of the wave packet.
It follows that
\be
\langle Q_{out}^2(jn) \rangle
 = \frac{1}{2}\left(\alpha_{jn;\,\tilde{\omega}}^* -\beta_{jn;\,\tilde{\omega}}\right)^2 M
  +\frac{1}{2}\left(\alpha_{jn;\,\tilde{\omega}}-\beta_{jn;\,\tilde{\omega}}^*\right)^2 M^*
+\left|\alpha_{jn;\,\tilde{\omega}}^*-\beta_{jn;\,\tilde{\omega}}\right|^2\, N
+\frac{1}{2}\,\int d\omega'  \left|\alpha_{jn;\,\omega'}^*-\beta_{jn;\,\omega'}\right|^2.
\ee
Continuing like in Sec.\ref{sec:planewave} we finally arrive at the relation \eqref{VoutvsVinwp} and
and then find parameters $\tau$ and $\bar{n}$ through $\mathbb{T}$ and $\mathbb{N}$ (actually we do not need to go through the pair $(\mathbb{T}_c,\mathbb{N}_c)$ in canonical form, since the parameters $\tau$ and $\bar{n}$ are invariants under symplectic
transformations \cite{CGH:2006}).


\begin{acknowledgments}
This work has been supported by FQXi under the programme ``Physics of Observer 2016".
\end{acknowledgments}


\section*{Author Contribution Statement}
SM posed the problem and wrote the main manuscript text, while GG did most of the calculations. Both authors reviewed the manuscript.

\section*{Additional Information}
The authors have no competing financial interest in relation to the present work.

\end{document}